\documentstyle[aps,prb,manuscript]{revtex}

\begin{document}

\draft

\title{Nonlinear resonance reflection from and transmission
through a dense glassy system built up of oriented linear Frenkel
chains: Two-level model}

\author{E. Conejero Jarque}

\address{Departamento de F\'{\i}sica Aplicada, Universidad
de Salamanca, E-37008 Salamanca, Spain}

\author{V. A. Malyshev}

\address{National Research Center "Vavilov State Optical
Institute", Birzhevaya Liniya 12, 199034 Saint-Petersburg, Russia}

\date{\today}

\maketitle

\begin{abstract}

A theoretical study of the resonance optical response of
assemblies of oriented short (as compared to an optical
wavelength) linear Frenkel chains is carried out. Despite the fact
that the energy spectrum of a single chain is composed of the
bands of Frenkel exciton states, a two-level model is used to
describe the optical response of a single linear chain. We account
for only the (on-resonance) optical transition between the ground
state and the state of the one-exciton band bottom as having the
dominating oscillator strength as compared to the other states of
the one-exciton manifold. The (off-resonance) process of creation
of two excitons per chain is neglected because it requires a
higher excitation frequency due to the quasi-fermionic nature of
one-dimensional Frenkel excitons. A distribution of linear chains
over length resulting in fluctuations of all exciton optical
parameters, such as the transition frequency and dipole moment as
well as the radiative rate, are taken explicitly into account.

We show that both transmittivity and reflectivity of the film may
behave in a bistable fashion, originated from saturation of the
nonlinear refraction index, and analyze how the effects found
depend on the film thickness and on the inhomogeneous width of the
exciton optical transition. Estimates of the driving parameters
show that films of oriented J-aggregates of polymethine dyes at
low temperatures seem to be suitable species for the experimental
verification of the behavior found.

\end{abstract}


\section{Introduction}
\label{intro}

Since pioneering works of Jelley~\cite{Jelley36} and
Scheibe~\cite{Scheibe37}, who discovered the phenomenon of linear
aggregation  of polymethine dye molecules in solution with rising
their volume concentration and, as a result, drastic changes of
their optical properties (the appearance of a red-shifted narrow
band, called now J-band), these species have been the subject of
unremitting interest of reseachers. In the beginning of the
nineties, Wiersma and co-workers~\cite{deBoer90,Fidder90,Fidder91}
demonstrated the possibility of further narrowing of the J-band as
well as shortening of its emission lifetime by an order of
magnitude via cooling to the temperature of liquid helium. At
present, it is widely accepted that the unusual properties of such
systems originate from the fact that their optically active states
are Frenkel exciton states (see for a comprehensive review
Refs.\onlinecite{Spano94} and~\onlinecite{Kobayashi96}).

In recent time, a considerable attention has been drawn to the
problem of strong coupling of organic polymers and molecular
aggregates to resonant radiation. Room temperature spectral
narrowing of emission~\cite{Tessler96,Hide96} and cooperative
emission~\cite{Frolov97,Frolov98} in $\pi$-conjugated polymer thin
films, superradiant lasing from the J-aggregated cyanine dye
molecules adsorbed onto colloidal silica and
silver~\cite{Akins97,Akins98} as well as the room temperature
polariton emission from strongly coupled organic semiconductor
microcavities~\cite{Lidzey99} have been reported. All these
effects unambiguously mean a collectivization of polymers and
J-aggregates via the emission field and are of great importance
from the viewpoint of laser applications (see
Refs.~\onlinecite{Tessler99,Kranzelbinder00,McGehee00} for
reviews). On the other hand, the conditions of such a strong
coupling are those necessary for manifestation of a bistable
behavior of a collection of homogeneously broadened two-level
systems.~\cite{Zakharov88,Basharov88,Benedict90,Benedict91,%
Benedict96,Logvin93} Much efforts have been undertaken to this
problem with its projection to J-aggregated assemblies. The
conditions for the bistable optical response to observe from an
individual
J-aggregate~\cite{Gusev92,Malyshev96,Malyshev97,Malyshev98a} and
even of a dimer~\cite{Gusev92,Heber87,Bodenschatz96,Malyshev98b}
have
been analized. Bistability discussed in Refs.~\onlinecite{Gusev92,%
Malyshev96,Malyshev97,Malyshev98a,Malyshev98b} is attributed to
{\it an individual} aggregate and consists of an abrupt switching
of the aggregate population from a low level to a higher one as
the pump intensity rises. Consequently, the reflectivity
(transmittivity) of a macroscopic ensemble of such species may be
switched off (on) in a step-wise way, allowing thus creation of an
all-optical bistable element.

The effect we are speaking about originates from the dynamical
resonance frequency shift dependent on the population of the
system.
However, as it has been shown in Refs.~\onlinecite{Malyshev96,%
Malyshev98a} and~\onlinecite{Malyshev98b}, an aggregate of a
length smaller than the emission wavelength does not display any
bistable behavior. This finding makes the above mechanism of the
aggregate bistability to be hardly experimentally verified.
Indeed, despite an aggregate may incorporate to itself thousands
of molecules and thus may have a length larger than the emission
wavelength, in reality, only a subsystem of segments, consisting
of a portion of the aggregate (the so-called coherently bound
molecules), responds to the action of a resonant external
field.~\cite{Knapp84} Because of a disorder (of static and thermal
nature) of the surroundings, the exciton coherence length, being
the physical length of an aggregate in the absence of disorder, is
reduced to a size $N^*$ dependent on the degree of disorder. For
the real conditions, the number $N^*$ is usually smaller than the
emission wavelength counted in the lattice unit. At room
temperature, $N^*$ is determined basically by dephasing
(originated from thermal fluctuations of molecular positions) and
has an order of magnitude of several lattice
units.~\cite{Bogdanov91,Wang91} On decreasing the temperature,
disorder tends to be of static nature. It is determined by static
fluctuations of the aggregate structure as well as the
surroundings and becomes dominating as the temperature approaches
zero. The static disorder results in Anderson localization of the
exciton within an aggregate segment with the typical size of the
order
of several tens of lattice units.~\cite{deBoer90,Fidder90,%
Fidder91,Minoshima94,Potma98} The number of molecules within the
localization length plays now the role of $N^*$ and, in fact,
represents a higher limit for the number of coherently bound
molecules. Further increasing $N^*$ would be possible by means of
reducing the degree of static disorder that is generally out of
control.

In this connection, in Refs.~\onlinecite{Malyshev99a} the question
was risen whether {\it an ensemble} of localized exciton states
may behave in a bistable fashion. It was used the fact that the
exciton states, located within the same localization domain and
being active in the optical response, form a local energy
structure with a few levels (two or three) similar to the one
existing on a regular chain with length $N \sim
N^*$.~\cite{Malyshev91,Malyshev95a} In other words, an aggregated
sample can be modelled by an ensemble of short (compared to the
emission wavelength) linear Frenkel chains stochastically
distributed over their lengths $N$. It was shown that an ultrathin
film with a thickness much smaller than the emission wavelength
(which makes possible to use the mean-field approach) manifested
bistability of the resonance optical response.

In this paper, we follow the conjecture proposed in
Refs.~\onlinecite{Malyshev99a} and generalize the results reported
there to film thicknesses of the order of or larger than the
emission wavelength, when the mean-field approach, being an
adequate approximation for thinner films, is no longer valid. This
paper is organized as follows. In Sec.~\ref{model}, we present the
model and mathematical formalism. Section~\ref{UTF} deals with an
analysis of the feasibility of a bistable response from an
ultrathin film (under the condition of validity of the mean field
approximation), taking into account the fluctuations of all
exciton optical parameters, such as the one-exciton transition
frequency and dipole moment as well as the one-exciton exciton
radiative rate.~\cite{note1} Further (Sec.~\ref{numerics}), we
present results of numerical simulations for thin films of
thickness larger than the emission wavelength and determine the
ranges of driving parameters, for which the film reflectivity and
transmittivity manifests bistability. In Sec.~\ref{Discuss}, we
make estimates for J-aggregates of PIC and show that critical
parameters for the occurence of such a behavior seems to be
achievable for thin films at low temperatures. Finally,
Sec.~\ref{concl} concludes the paper. Some preliminary results of
the present study have been reported in our recent
paper.~\cite{Malyshev00a}

\section{Description of the model}
\label{model}

An elementary object of an ensemble we will be dealing with
represents a short (with a length smaller than the emission
wavelength) ordered linear chain consisting of $N$ two-level
molecules (with $N$ much larger than unity). Due to the strong
intermolecular dipolar coupling, the optically active states are
the Frenkel exciton states rather than the states of individual
molecules. In the nearest-neighbour approximation, which we adopt
hereinafter, one-dimensional Frenkel excitons appear to be
non-interacting fermions~\cite{Chesnut63} so that any state with a
fixed number of excitons can be constructed as a Slater
determinant of one-exciton states
\begin{equation}
|k\rangle = \left({2\over N+1}\right)^{1/2} \sum_{n=1}^N \sin {\pi
kn\over N+1} |n\rangle, \ \ \ \ \ \ \ \ k = 1,2,....N \ ,
\label{k}
\end{equation}
where $|n\rangle$ is the ket-vector of the excited state of the
$n$th molecule. The energies of the one-dimensional exciton gas
may take the values $W = \sum_{k=1}^N n_k E_k$, where $n_k = 0,1$
is the occupation number of the $k$th one-exciton state and $E_k$
is the corresponding energy given by
\begin{equation}
E_k = \hbar \omega_{21} - 2U \cos{\pi k \over N+1} \ , \label{E}
\end{equation}
where $\omega_{21}$ is the transition frequency of an isolated
molecule and $U$ (chosen hereinafter to be positive) is the
magnitude of the nearest-neighbour hopping integral.

The optical transition from the  ground state of the chain to the
lowest state of the one exciton band $k=1$ has the dominating
oscillator strength ($81\%$ of the total one; see, for instance,
Ref.~\onlinecite{Spano94}). Furthermore, the transition betweeen
the bottoms of one-exciton and two-exciton bands is blue-shifted
as compared to the ground-state-to-one-exciton band bottom
transition by an energy $E_2-E_1 = 3\pi^2U/N^2$. The blue shift
originates of the fermionic nature of one-dimensional Frenkel
excitons and was experimentally verified for the first time in
Ref.~\onlinecite{Fidder93}. These two facts were put forward in
Ref.~\onlinecite{Malyshev99a} as a motivation to consider the
transition from the ground state to the bottom of the one-exciton
band as an isolated two-level transition. This assumption can
indeed be approved provided that the actual Rabi frequency of the
external field is smaller than the blue shift of one-to-two
exciton transitions. Further analysis of the transitions to higher
exciton manifolds showed that they did not lead to qualitative
changes of the behavior found within the framework of the
two-level model.~\cite{Glaeske01} Based on these findings and
taking into account that the spatial inhomogeneity of the field
and matter variables complicates the treatment drastically, we
will use in what follows the two-level approach to describe the
matter response.

Thus, we model the ensemble of linear Frenkel chains as that
comprised of inhomogeneously broadened (due to the fluctuations of
chain lengths) two-level systems, but with the optical
characteristics of the transition having all attributes of a
one-exciton transition. Note that neglecting the
one-to-two-exciton optical transitions means that no more than one
exciton per chain is created by the external field. Respectively,
such a channel of efficient exciton quenching as the intra-chain
exciton-exciton annihilation~\cite{Malyshev99b} plays no role
within the framework of the two-level approximation. In principle,
the excitonic annihilation can involve two excitons created on
different chains. As follows, however, from the results presented
in Ref.~\onlinecite{Malyshev99b}, the rate of such a process is
negligible under the restrictions of our model: low temperature
and $N \gg 1$.

We also assume that the transition dipole moments of all chains
are parallel to each other as well as  to the film
plane.~\cite{note2} With regard to the input field ${\bbox{\cal
E}_i}$, the on-resonance and normal incidence conditions are
chosen. The input field polarization can be set without loss of
generality to be directed along the transition dipole moment.
Then, all quantities can be considered as scalars.

Under the above limitations, the  time evolution of the film is
described in terms of the $2\times 2$ density matrix
$\rho_{\alpha\beta} \ (\alpha ,\beta = 1,2)$\ standing for
determination of the state of a chain of length $N$. The density
matrix equation together with the Maxwell equation for the total
field ${\cal E}$, including a secondary field produced by the
film, form the closed system of equations in the problem we are
dealing with. It reads
\begin{mathletters}
\label{1}
\begin{equation}
{\dot \rho}_{21} = - (i \omega + \Gamma) \rho_{21} - i \frac{d
{\cal E}}{\hbar} Z \ , \label{1_rho21}
\end{equation}
\begin{equation}
{\dot Z} = 2i \frac{d {\cal E}}{\hbar} [\rho_{12} - \rho_{21}] -
\Gamma_{1} (Z + 1)\ , \label{1_ZN}
\end{equation}
\begin{equation}
{\cal E}(x,t) = {\cal E}_i(x,t) -{2\pi\over c}\int_{0}^L dx^\prime
{\partial\over\partial t} {\cal
P}\left(x^\prime,t-{|x-x^\prime|\over c}\right)\ .
\label{1_maxwell_eq}
\end{equation}
\end{mathletters}
{}where the dots denote time derivatives; $Z = \rho_{22} -
\rho_{11}$; $\omega = \omega_{21} -
2(U/\hbar)\cos[\pi/(N+1)]\approx \omega_{21} - 2U/\hbar +
U\pi^2/\hbar N^2$ is the transition frequency for an individual
chain of size $N$; $d$ is the transition dipole moment of a chain
of size $N$ scaled as $d = d_0\sqrt{N}$\ , where $d_0$ is the
transition dipole moment for an isolated molecule; $\Gamma_{1}$ is
the spontaneous emission constant of the optically active
one-exciton state: $\Gamma_{1} = \gamma_0 N$ with $\gamma_0$ being
the analogous constant for an isolated molecule (for the sake of
simplicity, we have replaced the numerical factor $8/\pi^2$ in the
expression for $d$ and $\Gamma_{1}$ by unity); $\Gamma =
\Gamma_{1}/2 + \Gamma_2$ is the dephasing constant including the
contribution ($\Gamma_2$) not connected with the radiative
damping.

The last formula of the set (\ref{1}) is nothing else but the
integral form of the Maxwell equation for a film, in which $c$ and
$L$ stand for the speed of light and for the film thickness,
respectively, and ${\cal P}$ is the electric polarization:
\begin{equation}
{\cal P} = n_0 \Big \langle d (\rho_{21} +  \rho_{12}) \Big
\rangle \ , \label{P}
\end{equation}
where $n_0$ is the volume density of chains in the film and the
angle brackets denote averaging over the chain length distribution
with a probability distribution function $p(N)$: $\langle
...\rangle \equiv \sum_N p(N)...$.

As was pointed out in Refs.~\onlinecite{Benedict91,Benedict96,%
Benedict88,Malyshev95b,Malyshev97b,Jarque97,Manassah}, using the
integral wave equation to study the non-stationary nonlinear
response from a dense absorber has an obvious advantage: the
boundary conditions are not required within this framework.
Indeed, the reflected ${\cal E}_r = {\cal E}(0,t) - {\cal
E}_i(0,t)$ and transmitted ${\cal E}_{tr} = {\cal E}(L,t)$ fields
can be calculated obviously on the basis of
Eq.(\ref{1_maxwell_eq}) if the spatial distribution of the
electric polarization ${\cal P}$ is known. The latter, in turn, is
calculated from constituent equations~(\ref{1_rho21}) -
(\ref{1_ZN}) and~(\ref{P}). The background refraction index is not
included in Eq.(\ref{1_maxwell_eq}) because, in fact, this simply
results in constant renormalization.~\cite{Malyshev97b}

Rigorously speaking, for a dense system as that we are
considering, the deviation of the acting field from the average
(Maxwell) field may be significant. A simpleast way to account for
this difference is to add the (Lorentz-Lorenz) local-field
correction in the form $(4\pi/3){\cal P}$ to the Maxwell field
${\cal E}$ in the matter equations~(\ref{1_rho21})-(\ref{1_ZN})
(see, for instance, Ref.~\onlinecite{Benedict91}). Along this
paper, however, we will neglect the local field correction
justifying this approximation by our previous study: for that
arrangement of the incident fequency we are going to study (see
Sec.~\ref{Numerics}), this correction leads only to small
quantitative effects.~\cite{Malyshev97b}

Assume that the incident field has the form ${\cal E}_i =
E_i(t)\cos(\omega_i t - k_i x)$, where $\omega_i$ and $k_i =
\omega_i/c$ are the frequency and wavenumber, respectively, and
$E_i(t)$ is the amplitude, slowly varying in the scale of the
optical period $2\pi/\omega_i$. Thus, by passing to the rotating
frame by means of the representation $\rho_{21} = -(i/2) R
\exp(-i\omega_i t)$ and ${\cal E} = (1/2) E \exp(- i\omega_i t) +
c.c.$, with complex amplitudes $R$ and $E$ slowly varying in time,
and neglecting the counter-rotating terms, we obtain the set of
truncated equations
\begin{mathletters}
\label{2}
\begin{equation}
{\dot R} = - (i\Delta + \Gamma) R + {dE\over\hbar} Z \ ,
\label{2_RN}
\end{equation}
\begin{equation}
{\dot Z} = -{d\over 2\hbar} (E R^* + E^* R) - \Gamma_{1} (Z + 1) \
, \label{2_ZN}
\end{equation}
\begin{equation}
E(x,t) =  E_i\left(t-{x\over c}\right)e^{ik_ix} + 2\pi n_0
k_i\int^L _0 dx^\prime e^{ik_i|x-x^\prime |} \bigg \langle d R
\left(x^\prime,t-{|x-x^\prime |\over c}\right) \bigg \rangle\ ,
\label{2_maxwell_eq}
\end{equation}
\end{mathletters}
{}Here, the notation $\Delta = \omega - \omega_i$ is introduced.

The reflected and transmitted waves we are interested in are
simply given by the formulae:
\begin{mathletters}
\label{3}
\begin{equation}
E_r(t) = E(0,t) - E_0(t) =  2\pi k_i n_0 \int^L_0 dx e^{ik_ix}
\Big \langle d R\left(x,t- {x\over c}\right) \Big \rangle.
\label{ReflField}
\end{equation}
\begin{equation}
E_t(t) = E(L,t) = \left[ E_i(t) + 2\pi k_i n_0 \int^L_0 dx
e^{-ik_ix} \bigg \langle d R\left(x,t - {L-x \over c}\right) \bigg
\rangle \right] e^{ik_iL}\ . \label{TransmField}
\end{equation}
\end{mathletters}
As in our previous papers,~\cite{Malyshev95b,Malyshev97b,Jarque97}
we will neglect the retardation effects replacing $t-|x-x^\prime
|/c$ by $t$ since the film thickness of our interest is of the
order of a few vacuum wavelengths, $\lambda_i = 2\pi/k_i$.  Thus,
the actual passage time of the light through the film has the
order of magnitude of the optical period while all the
characteristic times of the problem discussed $(\Gamma,\ \hbar/d
E_i,\ \Delta^{-1})$ are much longer. One can, therefore, consider
the field as propagating instantaneously inside the film.

To perform numerical calculations, let us rewrite Eqs.\ (\ref{2})
in a dimensionless form using the dimensionless field amplitudes
$e ={\bar d}E/\hbar{\bar \Gamma}$ and $e_i ={\bar d}E_i/\hbar{\bar
\Gamma}$ as well as the dimensionless spatial and temporal
coordinates $\xi = k_ix$ and $\tau = {\bar\Gamma} t$, where ${\bar
d} = \langle d \rangle$ and ${\bar\Gamma} = \langle \Gamma
\rangle$ are the mean transition dipole moment and the mean
relaxation constant, respectively. One thus gets
\begin{mathletters}
\label{4}
\begin{equation}
{\dot R} = - (i\delta + \gamma) R + \mu e Z \ ,
\label{UdimBlochEqA}
\end{equation}
\begin{equation}
{\dot Z} = -{1\over 2}\mu (e R^* + e^* R) - \gamma_{1} (Z + 1) \ ,
\label{UdimBlochEqB}
\end{equation}
\begin{equation}
e(\xi,\tau) = e_i(\tau)e^{i\xi} + \Psi \int^{k_iL} _0 d\xi^\prime
e^{i|\xi -\xi^\prime |} \Big \langle \mu R(\xi^\prime,\tau) \Big
\rangle\ , \label{UdimIntWaveEq}
\end{equation}
\begin{equation}
e_r(\tau) = \Psi \int^{k_iL} _0 d\xi e^{i\xi} \Big \langle \mu
R(\xi,\tau) \Big \rangle\ , \label{e_r}
\end{equation}
\begin{equation}
e_t(\tau) = \left[e_i(\tau) + \Psi \int^{k_iL} _0 d\xi e^{-i\xi}
\Big \langle \mu R(\xi,\tau) \Big \rangle \right] e^{ik_iL} \ ,
\label{e_t}
\end{equation}
\end{mathletters}
{}where $\delta = \Delta/{\bar\Gamma}$\ , $\gamma =
\Gamma/{\bar\Gamma}$\ , $\mu = d/{\bar d}$\ ,  $\gamma_{1} =
\Gamma_{1}/{\bar\Gamma}$, and $\Psi = 2\pi d^2
n_0/\hbar{\bar\Gamma}$\ .

In our simulations, we choose as distribution function $p(N)$ a
Gaussian centered around ${\bar N}$ with standard deviation $a$
\begin{equation}
p(N) = {1 \over \sqrt{2\pi}a} \exp\left[-{(N - {\bar N})^2 \over
2a^2}\right]\ , \label{p}
\end{equation}
assuming $a < {\bar N}$. Under this assumption, it is easy to show
that the distribution function of the detuning $\delta$ also
presents a Gaussian centered at $\delta_0 = (\omega_{21} -
2U/\hbar + U\pi^2/\hbar {\bar N}^2- \omega)/ {\bar\Gamma}$ with
standard deviation $\sigma = 2\pi^2Ua/\hbar{\bar\Gamma} {\bar
N}^3$. The quantity $\sigma$ can be identified with the
inhomogeneous width of the exciton absorption line, so that the
limits $\sigma \ll 1$ and $\sigma \gg 1$ (or $2\pi^2Ua/\hbar{\bar
N}^3 \ll {\bar\Gamma}$ and $2\pi^2Ua/\hbar{\bar N}^3 \gg
{\bar\Gamma}$ in dimensional units) correspond to the cases of
dominating homogeneous and inhomogeneous broadening, respectively.

Equations~(\ref{4}) constitute the basis of our analysis of the
optical response from a thin film consisting of linear Frenkel
chains. We will be interested in the reflection and transmission
coefficients for amplitude, which are given by ${\cal R} =
\Big|[e(0,\tau) - e_i(0,\tau)]/e_i(0,\tau)\Big|$ and ${\cal T} =
\Big|e(k_iL,\tau)/e_i(k_iL,\tau)\Big|$.

\section{Ultrathin film, $L < \lambda^\prime$}
\label{UTF}

We turn first to the case of an ultrathin film ($L <
\lambda^\prime$ with $\lambda^\prime$ being the wavelength inside
the film) when the spatial dependence of $R$ and $Z$ can be
neglected and the exponentials in Eq.~(\ref{UdimIntWaveEq}) can be
replaced by unity inside the film. Then, equations for the fields
are simplified drastically:
\begin{mathletters}
\label{5}
\begin{equation}
e(\tau) = e_i(\tau) + \psi \big \langle \mu R(\tau) \big \rangle\
, \label{e}
\end{equation}
\begin{equation}
e_r(\tau) = \psi \big \langle \mu R(\tau) \big \rangle\ ,
\label{5e_r}
\end{equation}
\begin{equation}
e_t(\tau) = e_i(\tau) + \psi \big \langle \mu R(\tau) \big
\rangle\ , \label{5e_t}
\end{equation}
\end{mathletters}
where $\psi = \Psi k_iL$.

First of all, we are interested in the stationary states in which
the system can be found. This implies to look for steady-state
solutions to Eqs.~(\ref{4}), i.e., letting ${\dot R} = {\dot Z} =
0$. Under the simplification (\ref{5}), valid for an ultrathin
film, it is straightforward to arrive to a closed equation for the
transmission coefficient ${\cal T}$:
\begin{eqnarray}
{\cal T}^2\left[ \left( 1 + \psi \biggl\langle \mu^2 {\gamma
\over\delta^2 + \gamma^2 + \mu^2 e_i^2(\gamma / \gamma_{1}) {\cal
T}^2} \biggr\rangle \right)^2 \right. \nonumber\\ \nonumber\\
\left. + \psi^2 \biggl\langle \mu^2 {\delta \over \delta^2 +
\gamma^2 +  \mu^2 e_i^2 (\gamma /\gamma_{1}) {\cal T}^2}
\biggr\rangle^2 \right] = 1 \ . \label{TT}
\end{eqnarray}
Whenever ${\cal T}$ is found, the reflection coefficient ${\cal
R}$ can be expressed through ${\cal T}$ as follows:
\begin{eqnarray}
{\cal R}^2 = \psi^2 \left[\biggl\langle \mu^2 {\gamma
\over\delta^2 + \gamma^2 + \mu^2 e_i^2 (\gamma / \gamma_{1}) {\cal
T}^2} \biggr\rangle^2 \right. \nonumber\\ \nonumber\\ \left. +
\biggl\langle \mu^2 {\delta \over \delta^2 + \gamma^2 +  \mu^2
e_i^2 (\gamma /\gamma_{1}) {\cal T}^2} \biggr\rangle^2 \right]
{\cal T}^2 \ . \label{RR}
\end{eqnarray}
In absence of the chain length fluctuations, Eq.~(\ref{TT}) is of
third order with respect to ${\cal T}^2$ and thus may have three
real roots,~\cite{Malyshev99a} indicating the possibility of
bistable behavior of the system transmittivity. It occurs at $\psi
> \psi_c = 8$. Due to the relation~(\ref{RR}), the system
reflectivity is expected to behave in the same manner. The chain
length fluctuations make in general impossible to predict the
order of the resulting (after carrying out the averaging
procedure) transcendent equation.

In Ref.~\onlinecite{Malyshev99a}, Eq.~(\ref{TT}) was analyzed
under the assumption that the main effect of the length
fluctuations originates in the distribution of detuning $\delta$,
while fluctuations of the rest of parameters ($\mu,\ \gamma,\
\gamma_{1}$) are of no importance. Replacing then the distribution
over length, $p(N)$, by a distribution over detuning and taking
the latter in the form of a Lorentzian of width $g$, authors of
Ref.~\onlinecite{Malyshev99a} could evaluate integrals in
Eq.~(\ref{TT}) analytically and found that the resulting equation
gaves rise to a bistable behavior, even under the condition of
dominating inhomogeneous broadening ($g > \gamma$).

Here, we do not restrict ourselves to the above assumption and
carry out calculations, accounting for fluctuations of all
stochastic variables in Eq.~(\ref{TT}). We present the results for
a particular case of the incident frequency tuned up to the center
of the absorption band ($\delta_0 = 0$). Figure~\ref{Fig1} shows a
"phase" diagram of the system behavior in the space of parameters
($\sigma$, $\psi$) calculated on the basis of Eq.~(\ref{TT}). The
parameters of Gaussian $p(N)$ were chosen as ${\bar N} = 30$ and
$a = 9$. By the terms "one solution" and "three solutions" in
Fig.~\ref{Fig1}, we marked regions where Eq.~(\ref{TT}) had
single-valued and three-valued real solutions, respectively. The
solid curve, which separates these two regions, is nothing but the
critical value of $\psi$ for bistability to occur versus the
inhomogeneous width $\sigma = 2\pi^2Ua/\hbar{\bar\Gamma} {\bar
N}^3$. It starts from $\psi_c = 8$ at $\sigma = 0$, in full
correspondence with the earlier
findings,~\cite{Zakharov88,Basharov88,Malyshev99a} and then
gradually  goes up on increasing $\sigma$.

For comparison, we also depicted in Fig.~\ref{Fig1} the curve
(dashed) obtained under the approximations used in
Ref.~\onlinecite{Malyshev99a}, i.e., assuming the detuning
$\delta$ as being the unique stochastic variable and setting $\mu
= \gamma = 1, \gamma_1 = const$. As can be seen, there is almost
no difference between the solid and dashed curves. This means that
indeed, fluctuations of the detuning basically determine the
result of averaging in Eq.~(\ref{TT}), thus approving this
heuristic assumption used in Ref.~\onlinecite{Malyshev99a}.

Figure~\ref{Fig2} presents the typical examples of the input field
dependence of the reflection and  transmission coefficients for an
ultrathin film  ($k_iL = 0.1$) which were obtained by the
numerical solution of Eqs.~(\ref{4}) at adiabatic scanning of the
input field amplitude $e_i$ up and down. The results were obtained
chosing the following set of parameters: $\psi = 20$, ${\bar N} =
30$, $a = 9$, $\gamma_1  = 0.25 N / {\bar N}$, $\gamma  = 0.875 +
0.125 N / {\bar N}$. From this figure, one can conclude that, when
the inhomogeneous width $\sigma$ is small, the system shows a
stable hysteresis loop (optical hysteresis) both in reflectivity
and transmittivity, despite the fact that fluctuations of all
parameters are taken into account (note that the hysteresis of
transmittivity in Ref.~\onlinecite{Malyshev99a} was calculated
only for the particular case of absence of the chain length
fluctuations). However, as $\sigma$ grows, the hysteresis cycle
disappears and only a one-valued monotonic response exists.

In order to gain insight into the time required for the output
signal to approach its stationary value, we have numerically
solved Eqs.~(\ref{4}) at a fixed amplitude of the input field,
$e_i$, and its further step-wise switching to another value.
Figure~\ref{Fig3} shows an example of such calculations for the
same set of parameters as in Fig.~\ref{Fig2}, setting $\sigma =
0.25$. We start with a non-saturating incident field amplitude
$e_i = 3$, for which the reflection (transmission) coefficient is
high (low), and wait for a stationary value of the latter. At some
instant of time (in particular, at $\tau = {\bar\Gamma}t = 50$),
we switch step-wisely the incident field to a saturating value
$e_i = 6$, for which the reflection (transmission) coefficient is
low (high), and wait again for a stationary value of the latter.
The results of these calculation are depicted in Fig.~\ref{Fig3}
with solid lines. With dotted lines, we present the results of
analogous calculations obtained only by switching back the
incident field amplitude from a higher ($e_i = 6$) to a lower
($e_i = 3$) value.

From these data, it can be claimed first that the transient time
depends on whether the incident field amplitude is switched up or
down. Second, the transient time is of the order of a few units of
the population relaxation time ${\bar\Gamma}_1^{-1}$. Indeed, in
units of ${\bar\Gamma}^{-1}$ this time interval is of the order of
10. Taking into account that ${\bar\Gamma}_1/{\bar\Gamma} = 0.25$
in this particular calculation, one arrives at the above stated
conclusion.

\section{Thick film, $L > \lambda^\prime$}
\label{numerics}

\subsection{Motivation}
\label{motiv}

In our previous studies of nonlinear resonant reflection from an
extended ($L > \lambda^\prime$) dense system of {\it homogeneously
broadened} two-level
molecules,~\cite{Malyshev95b,Malyshev97b,Jarque97} we found that
the reflectivity of such a system could show different regimes
(bistability and self-oscillations~\cite{Malyshev95b,Malyshev97b}
as well as chaos~\cite{Jarque97}) basically governed by the
parameter $\Psi$, instead of $\psi = \Psi kL$ for an ultrathin
film (see also Refs.~\onlinecite{Roso85,Roso90}). The physical
meaning of $\Psi$ is nothing but the halfwidth (in units of the
homogeneous width ${\bar\Gamma}$) of a gap in the spectrum of
field + molecules collective excitations - polaritons (polariton
splitting; see for more details the monograph by
Davydov~\onlinecite{Davydov71}). This gap appears in the vicinity
of the molecule-field resonance and requires the condition $\Psi
\gg 1$ to be fulfilled. The linear dielectric constant for
frequencies ranging within the gap is negative (the refractive
index is imaginary), implying total reflection of the light of
such frequencies.~\cite{Davydov71}

The physical origin of the effects reported in
Refs.~\onlinecite{Malyshev95b,Malyshev97b,Jarque97,Roso90} is
attributed to saturation of the refraction index by the field
acting within the polariton band gap. Let us recall briefly the
motivation rised in those papers. For an input field varying
slowly in the scale of the relaxation times $\Gamma^{-1},
\Gamma_1^{-1}$ (the case of our interest in the present study),
the medium adiabatically follows the field ($\dot R$ can be set to
zero). Under such conditions, Eqs.~(\ref{UdimBlochEqA})
and~(\ref{UdimIntWaveEq}) are equivalent to only one
equation:~\cite{Malyshev95b,Malyshev97b}
\begin{mathletters}
\label{7}
\begin{equation}
     {d^2 e\over d\xi^2} +  \varepsilon(|e|^2)e = 0
\label{DifWaveEq}\ ,
\end{equation}
\begin{equation}
     \varepsilon(|e|^2)  =1 + 2\Psi \left\langle \mu {i\gamma
     + \delta \over \gamma^2 + \delta^2 + \mu^2 |e|^2\gamma/
     \gamma_1}\right\rangle \ ,
\label{NonLinDielFunc}
\end{equation}
\end{mathletters}
The quantity $\varepsilon(|e|^2)$ is the field-dependent
dielectric function in which fluctuations of all parameters $(\mu,
\gamma, \gamma_1, \delta)$ are explicitly taken into account. The
limit of absence of the chain length fluctuations ($\mu = \gamma =
1$, $\gamma_1 = const$ and $\delta = const$) formally corresponds
to the case considered in
Refs.~\onlinecite{Malyshev95b,Malyshev97b,Jarque97,Roso90}:
\begin{equation}
     \varepsilon(|e|^2)  = 1 + 2\Psi {i + \delta \over
     1 + \delta^2 + |e|^2/\gamma_1} \ .
\label{OldCase}
\end{equation}
We assume now that $\Psi \gg 1$ and neglect for a moment the field
(linear case). Then, as can be seen from Eq.~(\ref{OldCase}), the
linear dielectric function $\varepsilon$ has a dominating negative
real part within a rather wide interval of changing the detuning,
in fact, from a few negative units to $-2\Psi$. A weak field
acting within this band will be totally reflected from the system
boundary. At higher amplitudes of the field, $\gamma
|e|^2/\gamma_1 \gg \Psi$, the dielectric function is saturated by
the acting field, i.e., goes to unity, producing conditions for
penetration of the field into the medium. The higher the field
amplitudes, the deeper the field penetrates into the medium. Now,
reflection occurs from a very narrow (due to a high magnitude of
$\Psi$) interface between the saturated and non-saturated
regions.~\cite{Roso85} This interface plays the role of an
"artificial" mirror giving rise to a feedback necessary for the
effects outlined above to build up.

It turned out that namely in the limit of larger $\Psi$, i.e., in
the presence of the polariton band gap, the system shows the above
mentioned peculiarities of nonlinear optical response that are
absent in the opposite case $\Psi \le 1$. When inhomogeneous
broadening dominates (the main range of our interest), one should
compare the inhomogeneous width $2\sigma$ with the width of the
gap $2\Psi$. A simple analysis of Eq.~(\ref{NonLinDielFunc})
reveals the clear result that a well-pronounced gap (where the
dielectric function has a dominating negative real part) develops
at $\Psi \gg \sigma$. An inhomogeneous broadening of the order of,
or higher than $2\Psi$ destroys the gap and, as a consequence, all
the nonlinear effects accompanying the presence of this gap.
Below, we show details of the influence of inhomogeneous
broadening on the nonlinear optical response of the film.

\subsection{Numerical simulations and discussions}
\label{Numerics}

All nonlinear features accompanying the presence of the polariton
gap were found to be stronger when the input field frequency lies
in the middle of the gap, i.e., at $\delta = -
\Psi$.~\cite{Malyshev95b,Malyshev97b,Jarque97,Roso90} Therefore,
when studying the nonlinear reflection from and transmission
through a film consisting of {\it inhomogeneously broadened}
two-level systems, we will also set $\delta = -\Psi$  and consider
$\Psi > \Psi_c = 4.66$ ($\Psi_c$ is the threshold for bistability
in reflection~\cite{Malyshev95b}) to get the optimal conditions
for the effects we are looking for. It is worth noting that, as
was found in Ref.~\onlinecite{Malyshev97b}, the local-field
correction at $\delta = -\Psi$ has a relatively small quantitative
effect. In particular, the threshold value $\Psi_c$ changes from
4.66 to 5.45 if the local-field correction is taken into account.

Figure~\ref{Fig4} presents examples of the input field dependence
of the reflection and transmission coefficients for a thin film of
thickness $L = \lambda_i$ depending on the inhomogeneous width
$\sigma = (2\pi^2a/{\bar N}^3)\cdot (U/\hbar{\bar\Gamma})$, where
${\bar N}$ and $a$ were fixed (${\bar N} = 30,\ a = 9$) and
$U/\hbar{\bar\Gamma}$ was varied. The other parameters were chosen
as follows: $\Psi = -\delta = 6$, $\gamma_1  = 0.25 N / {\bar N}$,
$\gamma = 0.875 + 0.125 N/{\bar N}$. The data were obtained at
adiabatic scanning of the input field amplitude $e_i$ up and down.
From this figure, one can conclude, first, that the system shows a
stable hysteresis loop both in reflectivity and in transmittivity
or, in other words, behaves in a bistable fashion until $\sigma <
\Psi$, i.e., until the inhomogeneous width approaches the
polariton splitting, in full correspondence with what we mentioned
above. The second observation, evident from Fig.~\ref{Fig4}, is
that the switching amplitudes of the input field do not depend
strongly on the inhomogeneous width (at a fixed magnitude of
$\Psi$).

In Fig.~\ref{Fig5}, we depicted spatial profiles of the amplitude
module of the field inside the film versus the input field
amplitude $e_i$. The calculations were performed for the same set
of parameters and conditions as those presented in
Fig.~\ref{Fig4}, only setting $\sigma=2$. Plots (a) and (b)
correspond to adiabatic scanning of $e_i$  up and down,
respectively. The darkness of a local differential domain is
proportional to the field amplitude module. Observing these plots,
we can claim the following. Despite the fact that the thickness of
the system does not exceed one wavelength in vacuum, $\lambda_i$,
the field inside the film is not uniform at any amplitude of the
input field. For not-saturating magnitudes of $e_i$, this is
simply explained by the fact that the (linear) refraction index in
this case (at $\delta = -\Psi$) $n\approx -1$, that, in turn,
gives rise to decreasing the field amplitude on a scale short
compared to $\lambda_i$.  At higher magnitudes of the input field,
ranging within the bistable interval (see Fig.~\ref{Fig4}), the
spatial inhomogeneity of the field demonstrates a global
character, originating in the interplay between the forward and
backward (in the present case, reflected from the back boundary of
the film) waves.

Comparison of the field profiles depicted in the plots ({\it a})
and ({\it b}) at a fixed magnitude of the input field reveals that
they differ one from the other. In other words, one may make a
statement about the existence of spatial hysteresis of the field
inside the film.

Figure~\ref{Fig6} sheds light on the transient time of approaching
the reflectivity (a) and transmittivity (b) their stationary
values when switching the incident field amplitude up (solid
lines) and down (dotted lines), similar to that presented in
Fig.~\ref{Fig3}. The calculations were done for the following
parameters: $\Psi = -\delta = 6$, ${\bar N} = 30$, $a = 9$,
$\sigma = 2$, $\gamma_1 = 0.25 N/ {\bar N}$, $\gamma = 0.875 +
0.125 N/{\bar N}$  As in the case of an ultrathin film, the
transient time has the order of several units of the population
relaxation time ${\bar\Gamma}_1^{-1}$.

In Fig.~\ref{Fig7} we present the data of similar calculations as
depicted in Fig.~\ref{Fig4} performed for thicker films: $L =
3\lambda_i$ and $L = 5\lambda_i$. As follows from these results,
the hysteresis in reflection remains to be very pronounced.
However, it tends to disappear in transmittivity. It is simply due
to the fact that in order to make transparent a film of higher
thickness, one needs to apply an input field of very high
amplitude, for which the system already does not manifest at all
bistability in transmission. By contrast, with regards to
reflection, the film behaves as a "semi-infinite" medium, for
which bistability in reflection may occur provided the parameter
$\Psi$ exceeds its critical value $\Psi_c =
4.66$.~\cite{Malyshev95b}. In Fig.~\ref{Fig8}, we depicted the
spatial profile of the field inside the film with thickness $L =
3\lambda_i$ for the conditions for bistability to occur. It is
clear from these pictures that the field is absolutely attenuated
within a depth a bit longer than $\lambda_i$ and, hence, the
transmittivity in this case is negligible.

As follows from our previos
study~\cite{Malyshev95b,Malyshev97b,Jarque97} of the nonlinear
reflection from a collection of dense {\it homogeneosly broadened}
two-level atoms, the system may show instabilities of different
types (self-oscillation and chaos) at higher values of the
polariton splitting $\Psi$. Figure~\ref{Fig9} presents an example
of such a behavior of the film reflectivity (self-oscillation) for
$\Psi = 10$ and the {\it inhomogeneous broadening} $\sigma = 2$.
The attempt to get a hysteresis loop of the reflection coefficient
reveals that the lower branch is unstable (see Fig.~\ref{Fig9}{\it
a}). The calculation done at a fixed amplitude of the input field
($e_i = 8.5$, above the switching threshold) shows that indeed,
the system has a regime of self-oscillations with a frequency of
the order of ${\bar\Gamma}/2$.

\section{Estimates of parameters}
\label{Discuss}

First, let us discuss the applicability of the two-level model
used in this Paper. One should compare the typical energy spacing
between the first and second exciton levels $(E_2 -
E_1)/\hbar{\bar\Gamma} \simeq 3\pi^2 (U/\hbar{\bar\Gamma})/{\bar
N}^2$ with the switching magnitude of the input field $e_i$. For
the parameters at hand, the former quantity ranges within the
interval $[3.3, 33]$ while the latter is approximately equal to
$3$. Therefore, we can conclude that our approach is correct and
there is no necessity to include the one-to-two exciton
transitions.

It is worthwhile to analyze the parameters of real systems in
order to get insight into feasibility of the bistable mechanism we
are dealing with. In this sense, the J-aggregates of PIC, as one
of the most studied species of the type we need, seem to be
suitable objects. The width of the red J-band of PIC-Br (centered
at $\lambda = 576.1$ nm), at low temperature, has an inhomogeneous
nature and an order of magnitude of $30 cm^{-1}$ (or $1 ps^{-1}$
in frequency units)~\cite{deBoer90,Fidder90,Fidder91}. The blue
shift of the transition from one-to-two exciton bands with respect
to that from the ground state to the one-exciton band for the red
J-band has the same order. The polariton splitting is $\Psi\Gamma
= 2\pi {\bar d}^2 n_0/\hbar = (3/16\pi^2)\gamma_0 {\bar
N}n_0\lambda^3$, so that taking $\gamma_0 = (1/3.7)ns^{-1}$\ and
${\bar N}n_0 = 10^{18} cm^{-3}$\ (an achievable concentration of
molecules before aggregation), we obtain $\Psi\Gamma \simeq
1ps^{-1}$, i.e., a value that exceeds twice the halfwidth of the
J-band $0.5 ps^{-1}$ ($\sigma\Gamma$ in our notation). Hence, for
the parameters used, we are under the conditions needed for
bistability to occur.

\section{Conclusions and remarks}
\label{concl}

In this Paper, we have numerically studied the optical bistable
response of a thin film with thickness of the order of and larger
than the emission wavelength, comprised of oriented linear Frenkel
chains. We have taken into account a distribution of chains over
length, resulting in an inhomogeneous broadening of the exciton
optical transition as well as in the distribution of the
transition dipole moment.

From our study, the following conclusions can be drawn:

(i) Within a certain range of driving parameters (resonance
detuning, polariton splitting, and inhomogeneous width), the films
of thickness of the order of the vacuum emission wavelength really
show a bistable behavior with respect to both the reflection and
transmission of the resonant light;

(ii) The bistability effect exists until the inhomogeneous width
of the transition approaches the polariton splitting, i.e., for a
fairly wide range of widths;

(iii) With increasing the film thichness, the bistability in the
transmitted signal tends to disappear while it remains in the
reflected signal;

(iv) At higher magnitude of the linear refraction index, the film
reflectivity shows instabilities wich are of self-oscillating
type;

(v) The parameters needed for the bistable behavior seem to be
achievable for a thin film of J-aggregates of polymethine dyes and
for some classes of conjugated polymers deposited onto a
dielectric substrate such as poly({\it p}-phenylene-vinylene)
derivatives.

The inhomogeneous broadening acts as a destructive factor.
Reducing any source of inhomogeneity is thus of great importance.
In this sense, thin films comprised of thiophene oligomers seem to
be very attractive objects, too, due to the possibility of precise
controlling the sizes of this type of
molecules.~\cite{Ostoja93,Servet93,Blinov95,Garnier98}

To conclude, it is to be noting that the authors of
Ref.~\onlinecite{Daehne99} reported a room-temperature formation
of polariton states in ordered cyanine dye films. The latter thus
can also be considered as a promising object from the viewpoint of
our findings.

\acknowledgments

V.\ A.\ M.\ acknowledges a partial support from INTAS (project No.
97-10434).  E. C. J. acknowledges support by the Spanish
Direcci\'on General de Ense\~nanza Superior e Investigaci\'on
Cient{\'\i}fica (grant PB98-0268) and by the Junta de Castilla y
Le\'on in collaboration with European Union, F. S. E. (grant
SA44/01).

\begin{figure}
\caption{ ($\sigma,\psi$)-map obtained by solving
Eq.~(\protect\ref{TT}). The solid line separates two regions
within which there are one and three real solutions of
Eq.~(\protect\ref{TT}) and thus represents the critical value of
$\psi$ for bistability to occur versus the inhomogeneous width
$\sigma = 2\pi^2Ua/\hbar{\bar\Gamma} {\bar N}^3$. The dashed line
is calculated under the assumption that $\delta$ is the only
stochastic parameter, as was done in
Ref.~\protect\onlinecite{Malyshev99a}.} \label{Fig1}
\end{figure}

\begin{figure}
\caption{Amplitude reflection (a) and transmission (b)
coefficients of an ultrathin film ($k_iL = 0.1$) calculated using
Eqs.~(\protect\ref{4}) with the input field amplitude $e_i = {\bar
d}E_i/\hbar{\bar\Gamma}$ scanned up and down for different values
of $\sigma = 2\pi^2Ua/\hbar {\bar\Gamma}{\bar N}^3$ where
$U/\hbar{\bar\Gamma}$ was varied. Other parameters were chosen as
follows: $\psi = 20$, ${\bar N} = 30$, $a = 9$, $\gamma_1 = 0.25 N
/ {\bar N}$, $\gamma  = 0.875 + 0.125 N / {\bar N} $. Time is in
units of ${\bar\Gamma}^{-1}$.} \label{Fig2}
\end{figure}

\begin{figure}
\caption{ Kinetics of approaching the reflection (a) and
transmission (b) coefficients their stationary values after a
sudden switching (at an instant  $\tau = {\bar\Gamma}t = 50$) of
the incident field amplitude. The calculations were done for the
parameters of Fig.~\protect\ref{Fig2} setting $\sigma = 0.25$. The
solid lines represent the results obtained when the input field
amplitude was switched from a low value $e_i = {\bar
d}E_i/\hbar{\bar\Gamma} = 3$ (a high reflection) to a higher value
$e_i = {\bar d}E_i/\hbar{\bar\Gamma} =  6$ (a low reflection). The
dotted lines represent the results for the back switching of the
input field amplitude, from $e_i = 6$ to $e_i = 3$. Time is in
units of ${\bar\Gamma}^{-1}$.} \label{Fig3}
\end{figure}

\begin{figure}
\caption{The amplitude reflection (${\cal R}$) and transmission
(${\cal T}$) coefficients of a film with thickness $L = \lambda_i
\ (k_iL = 2\pi)$  calculated at adiabatic scanning of the input
field amplitude $e_i = {\bar d}E_i/ \hbar{\bar\Gamma}$ up and down
for different values of the inhomogeneous width $\sigma =
2\pi^2Ua/\hbar{\bar\Gamma}{\bar N}^3$, where $U/\hbar{\bar\Gamma}$
was varied. Other parameters were chosen as follows: $\Psi = 2\pi
{\bar d}^2n_0/\hbar{\bar\Gamma} = -\delta=6$, $\gamma_1  = 0.25 N
/ {\bar N}$, $\gamma = 0.875 + 0.125 N / {\bar N}$. Time is in
units of ${\bar\Gamma}^{-1}$.} \label{Fig4}
\end{figure}

\begin{figure}
\caption{ Examples of the spatial profiles of the field amplitude
module inside the film corresponding to the case with $\sigma = 2$
in Fig.~\protect\ref{Fig4}. The graphs (a) and (b) show the inside
field profiles when the input field amplitude $e_i = {\bar d}E_i/
\hbar{\bar\Gamma}$ is scanned up and down, respectively. The
darkness of a local differential domain is proportional to the
module of the inside field amplitude, $|e| = {\bar
d}|E|/\hbar{\bar\Gamma}$.} \label{Fig5}
\end{figure}

\begin{figure}
\caption{ Kinetics of approaching the reflection (a) and
transmission (b) coefficients their stationary values, calculated
after a sudden switching (at an instant $\tau = {\bar\Gamma}t =
50$) of the input field amplitude from a value $e_i = {\bar
d}E_i/\hbar{\bar \Gamma} = 2$ below the switching point $e_i
\approx 2.9$ to a one $e_i = {\bar d}E_i/ \hbar{\bar\Gamma} = 3.5$
above the latter (solid line). By the dotted line, the back
switching is shown. The parameters are the same as in
Fig.~\protect\ref{Fig5}. Time is in units of ${\bar\Gamma}^{-1}$.}
\label{Fig6}
\end{figure}

\begin{figure}
\caption{ The same as in Fig.~\protect\ref{Fig4} (a) for the case
of $\sigma=2$ and several values of the slab thickness.}
\label{Fig7}
\end{figure}

\begin{figure}
\caption{ The same as in Fig.~\protect\ref{Fig5} for a slab of
thickness $L=3 \lambda_i$.} \label{Fig8}
\end{figure}

\begin{figure}
\caption{{a} - Optical hysteresis loop of the reflection
coefficient for a film of thickness $L = 2 \lambda_i$ revealing an
instability.  The calculation was done by means of
Eqs.~(\protect\ref{4}) at adiabatic scanning of the input field
amplitude $e_i = {\bar d}E_i/ \hbar{\bar\Gamma}$ up and down,
setting the following parameters: $\Psi = 10 = - \delta$, ${\bar
N} = 30, a = 9$, $\sigma = 2$, $\gamma_1  = 0.75 N/{\bar N}$,
$\gamma = 0.625 + 0.375 (N /{\bar N})$. {b} - Kinetics of the
reflection coefficient ${\cal R}$ calculated at a fixed amplitude
of the input field ($e_i = 8.5$) showing self-oscillations. The
set of parameters is the same as in plot {a}. Time is in units of
${\bar\Gamma}^{-1}$.} \label{Fig9}
\end{figure}


\begin{references}




\bibitem{Jelley36} E.\ E.\ Jelley, Nature (London) {\bf 138}, 1009 (1936).



\bibitem{Scheibe37} G.\ Scheibe, Angew.\ Chem.\ {\bf 50}, 51 (1937).



\bibitem{deBoer90}S. de Boer and D. A. Wiersma, Chem. Phys. Lett.
     {\bf 165}, 45 (1990).



\bibitem{Fidder90}H. Fidder, J. Knoester, and D. A. Wiersma, Chem.
     Phys. Lett. {\bf 171}, 529 (1990).



\bibitem{Fidder91}H. Fidder, J. Terpstra, and D. A. Wiersma, J.
     Chem. Phys. {\bf 94}, 6895 (1991).



\bibitem{Spano94}F. C. Spano and J. Knoester, in {\it Advances in
     Magnetic and Optical Resonance\/}, Vol. {\bf 18}, ed. W. S.
     Warren (Academic, New York, 1994), p. 117.



\bibitem{Kobayashi96}T. Kobayashi (Ed.), {\it J-aggregates\/},
     World Scientific, Singapur, 1996.



\bibitem{Tessler96} N. Tessler, G. J. Denton, and R. H. Friend,
         Nature (London) {\bf 382}, 695 (1996).



\bibitem{Hide96}F. Hide, B. J. Schwartz, M. A. Diaz-Garsia, and
         A. J. Heeger, Chem. Phys. Lett. {\bf 256}, 424 (1996).



\bibitem{Frolov97}S. V. Frolov, W. Gellermann, M. Ozaki, K. Yoshino,
         and Z. V. Vardeny, Phys. Rev. Lett. {\bf 78}, 729 (1997).



\bibitem{Frolov98}S. V. Frolov, Z. V. Vardeny, and K. Yoshino,
         Phys. Rev. B {\bf 57}, 9141 (1998).



\bibitem{Akins97}S. {\" O}z\c{c}elik  and D. L. Akins,
         Appl. Phys. Lett. {\bf 71}, 3057 (1997).



\bibitem{Akins98}S. {\" O}z\c{c}elik, I. {\" O}z\c{c}elik, and
     D. L. Akins, Appl. Phys. Lett. {\bf 73}, 1949 (1998).



\bibitem{Lidzey99}D. G. Lizdey, D. D. C. Bradley, T. Virgili,
     A. Armitage, and M. S. Scolnik, Phys. Rev. Lett. {\bf 82},
     3316 (1999).



\bibitem{Tessler99}N. Tessler,
     Adv. Mater. {\bf 11}, 363 (1999).


\bibitem{Kranzelbinder00}G. Kranzelbinder and G. Leising,
     Rep. Prog. Phys. {\bf 63}, 729 (2000).


\bibitem{McGehee00}M. D. McGehee and A. J. Heeger,
     Adv. Mater. {\bf 12}, 1655 (2000).



\bibitem{Zakharov88}S. M. Zakharov and E. A. Manykin, Poverkhnost'
         {\bf 2}, 137 (1988).



\bibitem{Basharov88}A. M. Basharov, Zh. Exp. Teor. Fiz. {\bf 94},
     12 (1988) [JETP {\bf 67}, 1741 (1988)].



\bibitem{Benedict90}M. G. Benedict, A. I. Zaitsev, V. A. Malyshev,
         and E. D. Trifonov, Opt. Spektrosk. {\bf 68}, 812 (1990)
     [Opt. Spectrosc. {\bf 68} 473 (1990)].



\bibitem{Benedict91}M. G. Benedict, V. A. Malyshev, E. D. Trifonov,
     and A. I. Zaitsev, Phys. Rev. A {\bf 43}, 3845 (1991).



\bibitem{Benedict96}M. G. Benedict, A. M. Ermolaev, V. A. Malyshev,
         I. V. Sokolov, and E. D. Trifonov, {\it Super-radiance:
         Multiatomic coherent emission\/} (Bristol and Philadelphia:
         Institut of Physics Publishing, 1996).



\bibitem{Logvin93}Yu. A. Logvin and A. M. Samson, Opt. Commun.
     {\bf 96}, 107 (1993).



\bibitem{Gusev92}V. V. Gusev, Adv. Mater. Opt. Electr. {\bf 1},
     235 (1992).



\bibitem{Malyshev96}V. Malyshev and P. Moreno, Phys. Rev. A {\bf 53},
     416 (1996).



\bibitem{Malyshev97}V. A. Malyshev, H. Glaeske and K.-H. Feller,
     Opt. Commun. {\bf 140}, 83 (1997);
     J. Lumin. {\bf 76\&77}, 445 (1998);



\bibitem{Malyshev98a}V. A. Malyshev, H. Glaeske and K.-H. Feller,
     Phys. Rev. A {\bf 58}, 670 (1998).



\bibitem{Heber87} J.\ Heber, Z. Phys. B {\bf 68}, 115 (1987).



\bibitem{Bodenschatz96} N.\ Bodenschatz and J.\ Heber, Phys. Rev. A
         {\bf 54}, 4428 (1996).



\bibitem{Malyshev98b}V. A. Malyshev, H. Glaeske and K.-H. Feller,
         Phys. Rev. A {\bf 58}, 1496 (1998).



\bibitem{Knapp84}E. W. Knapp, Chem. Phys. {\bf 85}, 73 (1984).


\bibitem{Bogdanov91}V. I. Bogdanov, E. N. Viktorova, S. V. Kulya,
     and A. S. Spiro, Pis'ma Zh. Eksp. Teor. Fiz. {\bf 53}, 100
     (1991)[JETP Lett. {\bf 53}, 105 (1991).



\bibitem{Wang91}Y. Wang, J. Opt. Soc. Am. B {\bf 8}, 981 (1991).






\bibitem{Minoshima94}K. Minoshima, M. Taiji, K. Misawa,
     T. Kobayashi, Chem. Phys. Lett. {\bf 218} 67 (1994).



\bibitem{Potma98}E. O. Potma and D. A. Wiersma, J. Chem. Phys.
     {\bf 108}, 4894 (1998).



\bibitem{Malyshev99a}V. A. Malyshev, H. Glaeske and K.-H. Feller,
     Opt. Commun. {\bf 169}, 177 (1999);
     J. Lumin. {\bf 83-84}, 291 (1999);
     J. Chem. Phys. {\bf 113}, 1170 (2000).



\bibitem{Malyshev91}V. A. Malyshev, Opt. Spektrosk. {\bf 71},873
     (1991) [Opt. Spectrosc. {\bf 71}, 505 (1991)];
     J. Lumin. {\bf 55}, 225 (1993).



\bibitem{Malyshev95a}V. Malyshev and P. Moreno, Phys. Rev. B {\bf 51},
     14 587 (1995).



\bibitem{note1} In Ref.~\protect\onlinecite{Malyshev99a}, only
         the fluctuations of the one-exciton transition frequency
         (inhomogeneous broadening) have been taken into account.



\bibitem{Malyshev00a}V. A. Malyshev and E. Conejero Jarque, Opt.
     Exp. {\bf 6}, 227 (2000).



\bibitem{Chesnut63}D. B. Chesnut and A. Suna, J. Chem. Phys. {\bf 39},
     146 (1963).



\bibitem{Fidder93}H. Fidder, J. Knoester and D. A. Wiersma, J.
     Chem. Phys. {\bf 98}, 6564 (1993).



\bibitem{Glaeske01} H. Glaeske, V. A. Malyshev, and K.-H. Feller,
     J. Chem. Phys. {\bf 114}, 1966 (2001).



\bibitem{Malyshev99b}V. A. Malyshev, H. Glaeske and K.-H. Feller,
     Chem. Phys. Lett. {\bf 305}, 117 (1999);
     Chem. Phys. {\bf 254}, 31 (2000).



\bibitem{note2}Such conditions are achievable for thin films
     prepared by the spin coating method~\protect\cite{Misawa93}.



\bibitem{Misawa93}K. Misawa, K. Minoshima, H. Ono, and T. Kobayashi,
     Appl. Phys. Lett. {\bf 63}, 577 (1993).



\bibitem{Benedict88}M. G. Benedict and E. D. Trifonov, Phys. Rev. A
     38 (1988) 2854.



\bibitem{Malyshev95b}V. A. Malyshev and E. Conejero Jarque,
     J. Opt. Soc. Am. B 12 (1995) 1868;
     J. Lumin. 72-74 (1997) 822; Opt. Spektr. 82 (1997) 630
     [Opt. Spectr. 82 (1997) 582].



\bibitem{Malyshev97b}V. A. Malyshev and E. Conejero Jarque,
     J. Opt. Soc. Am. B 14 (1997) 1167.



\bibitem{Jarque97}E. Conejero Jarque and V. A. Malyshev, Opt.
     Commun. 142 (1997) 66.


\bibitem{Manassah}J. T. Manassah and B. Gross, Opt. Commun. 131
     (1996) 408; {\it ibid.} 148 (1998) 404; {\it ibid.} 149 (1998)
     393; {\it ibid.} 155 (1998) 213.



\bibitem{Roso85} L.\ Roso-Franco, Phys. Rev. Lett. 55 (1985) 2149;
     J. Opt. Soc. Am. B 4 (1987) 1878.



\bibitem{Roso90} L.\ Roso-Franco and M.\ Ll.\ Pons, J.\ Mod.\ Opt.\
     37 (1990) 1645.



\bibitem{Davydov71}A.\ S.\ Davydov, Theory of molecular excitons,
         Plenum Press, New York, 1971.



\bibitem{Ostoja93}P. Ostoja, S. Guerri, S. Rossini, M. Servidory,
     C. Taliani, and R. Zamboni, Synth. Met. {\bf 54}, 447 (1993).



\bibitem{Servet93}B. Servet, S. Ries, M. Tritel, P. Alnot,
     G. Horowitz, and H. Garnier, Adv. Mater. {\bf 5}, 461 (1993).



\bibitem{Blinov95}L. M. Blinov, S. P. Palto, G. Ruani, C. Taliani,
     A. A. Tevosov, S. G. Yudin, and R. Zamboni, Chem Phys. Lett.
     {\bf 232}, 401 (1995).



\bibitem{Garnier98}F. Garnier, G. Horowitz, P. Valat, and F. Kouki,
     Appl. Phys. Lett. {\bf 72}, 2087 (1998).



\bibitem{Daehne99}L. Daehne and E. Biller, Phys. Chem. Chem. Phys.
     {\bf 1}, 1727 (1999).


\end{references}
\end{document}